\documentclass[a4paper,11pt]{article}
\usepackage{cite}

\usepackage[utf8]{inputenc}
\usepackage[T1,T2A]{fontenc}
\usepackage{amsmath,amsfonts,amssymb}

\usepackage{graphicx}
\usepackage{hyperref}

\numberwithin{equation}{section}

\begin{document}

\begin{titlepage}

\begin{center} {\LARGE \bf Inflaton freeze--out} \end{center}

\vspace{1cm}

\begin{center}
  {\bf Oleg Lebedev\(^a\), Thomas Nerdi\(^a\),
  Timofey Solomko\(^b\), Jong-Hyun Yoon\(^a\)}
\end{center}

\begin{center}
  \vspace*{0.15cm}
  \it{\({}^a\)Department of Physics and Helsinki Institute of Physics,\\
  Gustaf H\"allstr\"omin katu 2a, FI-00014 Helsinki, Finland}\\
  \vspace*{0.15cm}
  \it{\({}^b\)Saint Petersburg State University, 7/9 Universitetskaya nab.,\\
  St.Petersburg, 199034, Russia}
\end{center}

\vspace{2.5cm}

\begin{center} {\bf Abstract} \end{center}

\noindent  We study the possibility that, after inflation, the inflaton reaches thermal equilibrium with the Standard Model thermal bath
and eventually freezes--out in the non--relativistic regime. When the inflaton decay is the sole source of (non--thermal) dark matter, its relic density is automatically suppressed.
We delineate parameter space leading to the correct dark matter abundance. The model allows for a significant Higgs--inflaton coupling which may lead to invisible Higgs decay into inflaton
pairs at the LHC.

\end{titlepage}

\tableofcontents

\section{Introduction}

The Standard Model (SM) of particle physics is challenged by
the existence of dark matter (DM) and an inflationary paradigm.
One of the minimalistic options to address these cosmological issues is to extend the SM with just 2 degrees of freedom in the form of two real scalars.
One scalar would then be responsible for driving inflation, while the other  would be stable and play the role of dark matter.
The inflationary energy must subsequently be converted into SM radiation, which necessitates  a coupling between the inflaton and some SM fields.
The leading renormalizable couplings are provided by the ``Higgs portal'' \cite{Silveira:1985rk,Patt:2006fw},
\begin{equation}
\Delta V = {1\over 2} \lambda_{\phi h} \phi^2 H^\dagger H +  \sigma_{\phi h} \phi H^\dagger H\;,
\end{equation}
 where $\phi$ is the inflaton and $\lambda_{\phi h},\sigma_{\phi h}$ are some coupling constants. On general grounds, these interactions are expected to be
  responsible for reheating the Universe.  Analogous couplings can be written down for the dark matter field, which would lead to DM production directly by
  the inflaton.

 The absence of the direct DM detection signal motivates one to consider seriously the possibility that dark matter couples feebly to normal matter.
 It may have never been in thermal equilibrium and its current abundance could be determined directly by its coupling to the inflaton.
   This framework has been analyzed  in detail in \cite{Lebedev:2021tas} and reviewed  in \cite{Lebedev:2021xey}.
   In our work, we extend the previous studies by considering inflaton thermalization due to its interaction with the Higgs field and subsequent inflaton freeze--out.
   This suppresses the inflaton energy density compared to that of the SM thermal bath. If inflaton decay is the only source of non--thermal dark matter, the relic abundance of the latter
   will consequently be suppressed, as required by observations.
 In what follows, we discuss the technical aspects of this mechanism and delineate parameter space leading to the correct DM abundance.

\section{Higgs portal framework}

The minimal Standard Model extension that accommodates dark matter and inflation includes
2 real scalars,  $\phi$ (inflaton) and $s$ (dark matter).\footnote{It is possible that the inflaton also plays the role of dark matter \cite{Liddle:2006qz,Lerner:2009xg},
yet the minimal option is strongly constrained \cite{Lebedev:2021zdh}.}
This framework is reviewed in \cite{Lebedev:2021xey}.
 The $only$ renormalizable inflaton  couplings to the Standard Model are
 \begin{equation}
 V_{\phi h} = {1\over 4} \lambda_{\phi h} \phi^2 h^2 + {1\over 2} \sigma_{\phi h} \phi h^2 \;,
 \end{equation}
 where we have assumed the unitary gauge for the Higgs field $H= (0,h/\sqrt{2})^{\rm T}$.
 The inflaton mass is denoted by $m_\phi$ and $\phi$ is taken to have a zero VEV.
The  DM couplings to the inflaton
 are  given by
  \begin{equation}
 V_{\phi s} = {1\over 4} \lambda_{\phi s} \phi^2 s^2 + {1\over 2} \sigma_{\phi s} \phi s^2 \;,
 \end{equation}
 where a stabilizing $Z_2$ symmetry $s \rightarrow -s$ has been imposed.
 The DM mass is denoted by $m_s$.
In what follows, we study the possibility that
  dark matter  is {\it non--thermal} and the
    Higgs--DM interaction
    \begin{equation}
    V_{sh} = {1\over 4} \lambda_{sh}h^2 s^2
    \end{equation}
     is negligible, $\lambda_{sh} \rightarrow 0$.

We focus on the range of $\lambda_{\phi h}$ values leading to efficient inflaton--Higgs scattering and eventual thermalization of the inflaton--Higgs system.
As we show later, this sets the lower bound
\begin{equation}
\lambda_{\phi h} \gtrsim 10^{-8}-10^{-7} \;.
\end{equation}
 While the inflaton may be lighter or heavier than the Higgs, we require
\begin{equation}
m_\phi > 2 m_s \;,
\end{equation}
such that perturbative decay $\phi \rightarrow ss$ is allowed.  The trilinear couplings $\sigma_{\phi s}, \sigma_{\phi h}$ are assumed to be sufficiently small so that
they do not affect the inflaton thermodynamics until $\phi$ decays at late times.

After inflation, the Higgs quanta can be copiously produced via parametric resonance  \cite{Kofman:1994rk,Kofman:1997yn} induced by the coupling $\lambda_{\phi h}$. Following rescattering and thermalization,
the inflaton--Higgs system remains in thermal equilibrium until the inflaton freezes--out.
 If this occurs in the non--relativistic regime, its energy density is suppressed compared to that of the SM bath. Then,  the decay $\phi \rightarrow ss$ generates
 dark matter whose abundance is automatically small, in accordance with observations. This mechanism is reminiscent of the super--WIMP idea
 put forth in \cite{Covi:1999ty,Feng:2003uy}.

Clearly, for this scenario to work, the couplings have to be in a specific range. The different options are summarized in Table 1.
In particular, in order to eliminate additional  sources of dark matter such as freeze--in production, we take $\lambda_{ sh}\ll 10^{-11}$ \cite{Lebedev:2019ton}. Further,
if $\lambda_{\phi s} \gtrsim \lambda_{\phi h}$ and $\lambda_{\phi h}$ is large enough for thermalization, then dark matter also reaches thermal equilibrium with the inflaton.
Since it does not have an efficient annihilation channel for $m_\phi > m_s$, its abundance is bounded from below, roughly by the inverse of the number of degrees of freedom (see e.g. \cite{Lebedev:2021tas}).
 This makes dark matter overabundant, thus we require  $\lambda_{\phi s} \ll \lambda_{\phi h}$. Finally, the relation between the trilinear couplings $\sigma_{\phi s}$ and $ \sigma_{\phi h}$
 affects the efficiency of the SM state production in late inflaton decay. If $\phi \rightarrow {\rm SM }$ is non--negligible, the inflaton lifetime must be below ${\cal O}(0.1\, {\rm sec})$
 in order not to spoil the standard  nucleosynthesis.

\begin{center}
\begin{tabular}{| c |  c |   }
\hline
 coupling regime& feature  \\
 \hline
$\lambda_{ sh } \ll 10^{-11}$  & present model   \\
$\lambda_{ sh } \gtrsim 10^{-11}$  & DM freeze--in or freeze--out   \\
 $\lambda_{\phi s} \gtrsim \lambda_{\phi h}$ & too much DM  \\
$\lambda_{\phi s} \ll \lambda_{\phi h}$  & present model   \\
$\sigma_{\phi s}  \gg \sigma_{\phi h}$ & present model (no BBN constraint) \\
$\sigma_{\phi s} \lesssim  \sigma_{\phi h}$ & present model (with BBN constraint) \\
 \hline
\end{tabular}
\\ \ \\ Table 1:  Coupling regimes in the Higgs portal model.
\end{center}

\section{Motivation: inflation driven by a non--minimal scalar--curvature coupling}

The main premise of our work is that the inflaton reaches thermal equilibrium with the Standard Model thermal bath. Clearly, it requires a sufficiently large coupling between the two. This may be problematic since such a coupling generally induces a large loop correction  to the inflaton potential thereby spoiling its flatness.
However,
 in a class of models
based on a non--minimal scalar coupling to gravity,  the inflaton self--interaction can be significant and the loop corrections small compared to the tree level value.
Below we describe the main features of such models.

A simple and viable inflationary model is based on the action \cite{Bezrukov:2007ep}
\begin{equation}
{\cal L}_{J} = \sqrt{-\hat g} \left(   -{1\over 2}  \Omega  \hat R \,
 +  {1\over 2 } \, \partial_\mu \phi \partial^\mu \phi     - {V(\phi)  }\right) \;,
\label{L-J}
\end{equation}
with
\begin{equation}
 \Omega = 1 + \xi_\phi \phi^2 ~~,~~V(\phi) =   {1\over 4} \lambda_\phi \phi^4  +
  {1\over 2} m_\phi^2 \phi^2   \;.
\label{potential}
\end{equation}
Here we use the Planck units
\begin{equation}
M_{\rm Pl}=1 \;,
\end{equation}
where $M_{\rm Pl}$ is the reduced Planck mass;
$m_\phi \ll 1$;
$\hat g^{\mu \nu}$ denotes the Jordan frame metric and $\hat R$ is the corresponding scalar curvature.
The transition to the Einstein frame is accomplished by the metric rescaling
\begin{equation}
 g_{\mu\nu} = \Omega \, \hat g_{\mu\nu}   \;,
\end{equation}
such that the curvature based on the metric $g^{\mu \nu}$ appears in the Lagrangian with the canonical coefficient $-1/2$.
At large field values, $\Omega \simeq \xi_\phi \phi^2$ and the canonically normalized inflaton becomes \cite{Bezrukov:2007ep}
\begin{equation}
\chi= \sqrt{3\over 2} \ln (  \xi_\phi \phi^2) \;,
\end{equation}
with the potential
 \begin{equation}
V_E = {\lambda_{\phi} \over 4 \xi_\phi^2} \left( 1+  \exp \left(     - {2 \gamma \chi \over \sqrt{6}}  \right) \right)^{-2} \;,
  \label{VE1}
 \end{equation}
 where
  \begin{equation}
  \gamma = \sqrt{6 \xi_\phi \over 6\xi_\phi +1  } \;.
 \end{equation}
This inflaton potential is well consistent with the inflationary data. The COBE normalization requires \cite{Lebedev:2021xey}
\begin{equation}
 {\lambda_{\phi} \over 4 \xi_\phi^2} = 4\times 10^{-7} \, {1\over \gamma^2 N^2 } \;,
 \label{cobe}
 \end{equation}
where $N=50...60 $ is the number of inflationary $e$--folds. For $\xi_\phi \gtrsim 1/6$, this implies $ {\lambda_{\phi} \over 4 \xi_\phi^2} \sim 10^{-10}$.
The spectral index $n$ and the tensor-to-scalar ratio $r$ are given by
   \begin{eqnarray}
&&     n = 1-6 \epsilon + 2 \eta \simeq 1-{2\over N}  - {9\over 2 \gamma^2 N^2 }  \;, \nonumber\\
&&  r =16 \epsilon \simeq {12\over \gamma^2 N^2}  \;.
\label{n-r}
      \end{eqnarray}
These predictions fit  the PLANCK data  very well \cite{Planck:2018jri}.

After inflation, the inflaton starts oscillating with a decreasing amplitude and, for $\xi_\phi \phi^2 <1$,  the field $\phi$ becomes a canonically normalized  variable with the
potential (\ref{potential}). The inflaton can be taken to be light enough such that the potential is dominated by the quartic term. Oscillations in this potential
can lead to efficient particle production and, eventually,  reheating.

The model parameters are subject to a unitarity constraint. The non--minimal scalar coupling to gravity corresponds to a dimension--5 operator, which implies that the theory is valid up to a cutoff
\cite{Burgess:2009ea,Barbon:2009ya}
\begin{equation}
\Lambda \sim {1\over \xi_\phi} \;.
\end{equation}
The energy density during inflation,  $(\lambda_\phi/ 4\xi_\phi^2)^{1/4}$, should be below the cutoff. Combining this condition with the COBE normalization (\ref{cobe}),  for $\gamma \sim 1$ one finds
\begin{equation}
 \lambda_{\phi} (H)\lesssim 4\times 10^{-5}
 \label{uni-bound}
\end{equation}
  and $\xi_\phi (H) \lesssim 300$. Here $\lambda_{\phi} (H)$ is the running coupling evaluated at the Hubble scale $H$.
Thus, the inflaton self--coupling cannot be too strong. In turn, this implies that the loop corrections to this coupling cannot be too large either.
In particular, if one introduces the Higgs portal coupling
\begin{equation}
\Delta V = {1\over 4} \lambda_{\phi h} \phi^2 h^2 \;,
\end{equation}
where $h$ is the Higgs field in the unitary gauge, the resulting radiative correction to $\lambda_\phi$ is of order
$  \lambda_{\phi h}^2/(8 \pi^2)$, up to the logarithm of the renormalization group  scales ratio. The bound (\ref{uni-bound}) then implies
 \begin{equation}
\lambda_{\phi h } \lesssim 10^{-2} \;.
\end{equation}
This constraint is loose enough to allow for thermalization of the inflaton--Higgs system without inducing too large a correction to the inflaton potential.

The constraints relax further if one employs the {\it Palatini} \cite{Bauer:2008zj} instead of metric formalism. That is, in addition to the metric $g_{\mu \nu}$, one introduces the connection degrees of freedom $\Gamma^\lambda_{\mu\nu}$. In this case, the curvature $R_{\mu\nu}$ is a function of the connection only. Eliminating $\Gamma^\lambda_{\mu\nu}$ via their equations of motion, one finds a theory similar to the one described above
albeit with some important modifications. In particular, the canonically normalized inflaton $\chi$ is defined by the relation $\phi = 1/\sqrt{\xi_\phi} \, \sinh (\sqrt{\xi_\phi} \chi)$ such that the unitarity cutoff becomes
\cite{Bauer:2010jg}
\begin{equation}
\Lambda_{\rm Pal} \sim {1\over \sqrt{\xi_\phi}} \;.
\end{equation}
The energy density during inflation remains the same,
so unitarity is preserved as long as the system remains perturbative,
\begin{equation}
\lambda_\phi ~,~ \lambda_{\phi h} \lesssim {\cal O}(1) \;.
\end{equation}
As before, the Higgs--induced correction to the inflaton potential is small as long as $\lambda_\phi \gg \lambda_{\phi h}^2/(8 \pi^2)$.
A somewhat uncomfortable aspect of this approach is that the inflationary data
 require an extremely large $\xi_\phi \sim 10^{10} \lambda_\phi$.

The above examples show that there exist classes of inflationary models in which the inflaton--Higgs coupling can be quite large without spoiling the flatness of the inflaton potential.
Depending on the inflaton mass,
couplings of this size often suffice to bring the system to thermal equilibrium such that during reheating the inflaton shares a thermal bath with the Standard Model states. This question will be considered in more detail in the next section.

\section{Thermalization constraint}

The inflaton--Higgs system reaches thermal equilibrium for $\lambda_{\phi h}$ above a certain $m_\phi$--dependent value. The main relevant processes are
\begin{equation}
\phi \phi \leftrightarrow h_i h_i ~~,~~ \phi \phi \leftrightarrow h\;,
\end{equation}
where $h_i$ represents 4 Higgs degrees of freedom at high energies, while $h$ stands for a single Higgs d.o.f. at low energies. If the rate of these processes is above the Hubble rate $H$,
the Higgses are copiously produced and
thermalization sets in.
Let us consider these processes separately.

\subsection{\texorpdfstring{$ \phi \phi \rightarrow h_i h_i$}{phi phi -> hi hi}}
Comparing the corresponding terms in the Boltzmann equation
for the inflaton number density $n_\phi$, the thermalization condition can be formulated as
\begin{equation}
3n_\phi H < 2 \Gamma(\phi \phi \rightarrow h_i h_i )\;,
\end{equation}
where $ \Gamma(\phi \phi \rightarrow h_i h_i )$ is the reaction rate per unit volume.

\begin{figure}[t]
\centering{
\includegraphics[width=0.49\textwidth]{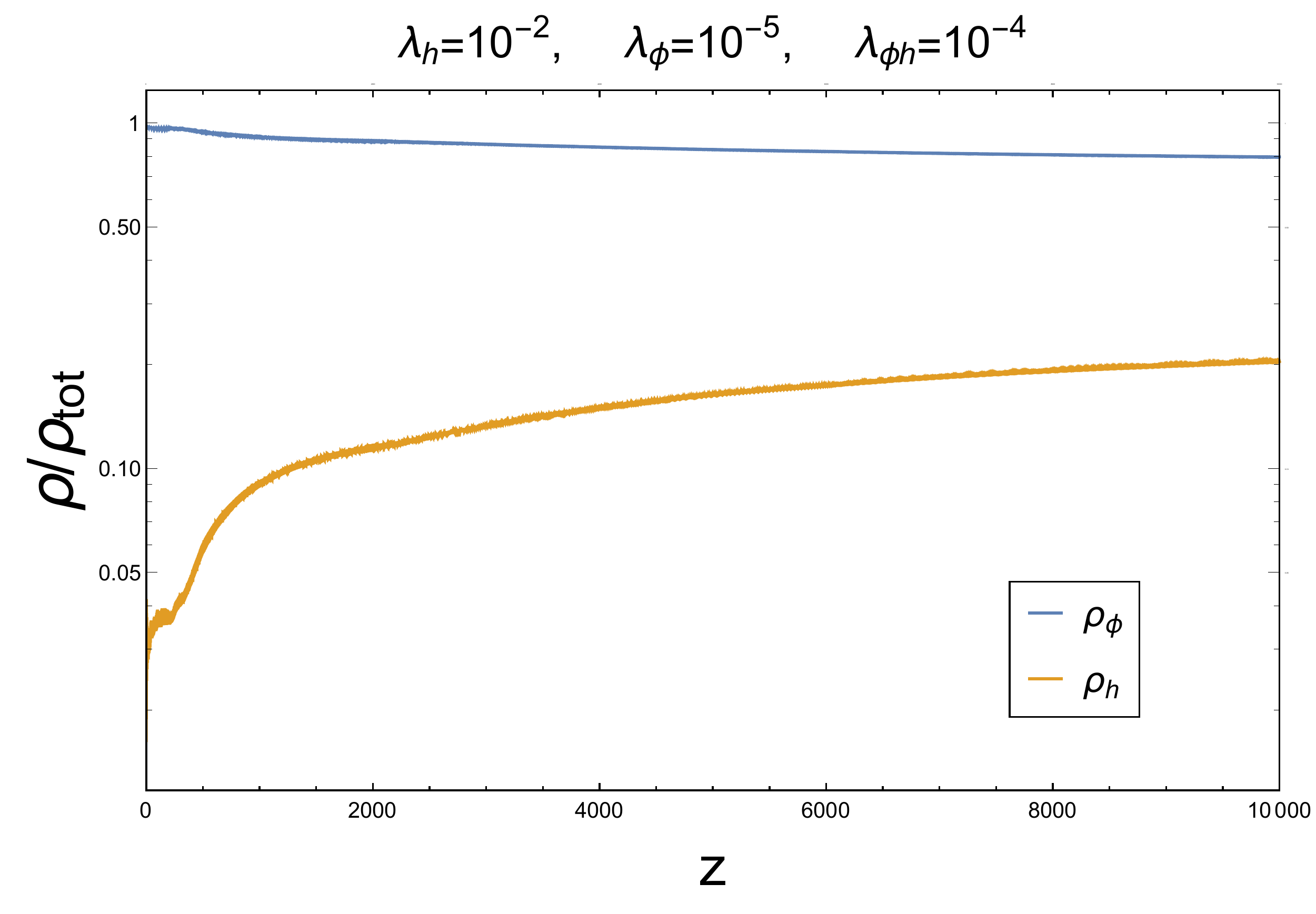}
\includegraphics[width=0.49\textwidth]{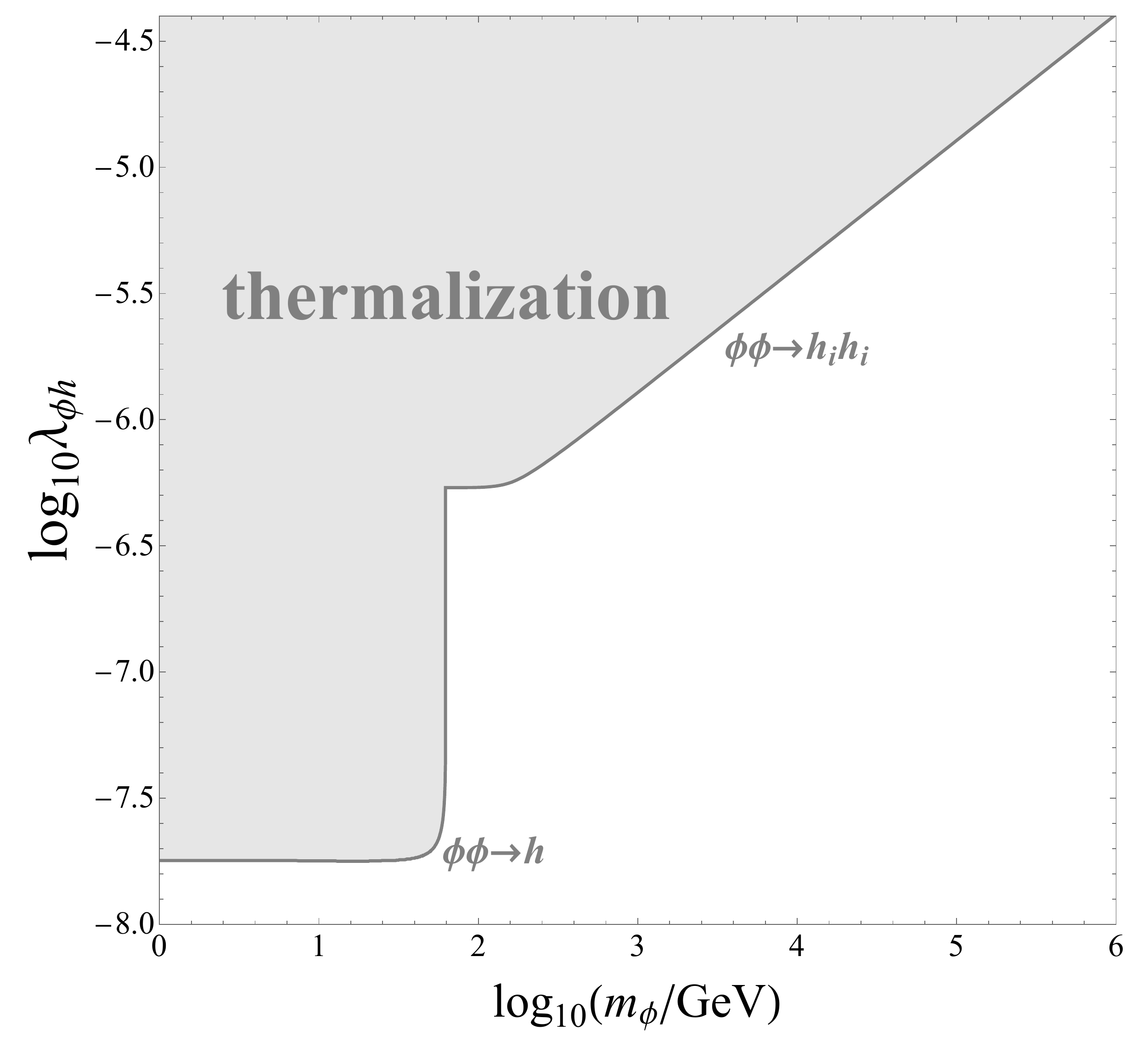}
}
\caption{ \label{therm-fig}
{\it Left:} fraction of the energy density carried by the inflaton  and 4 Higgs d.o.f. as a function of conformal time $z$.
The simulation is performed  with LATTICEEASY \cite{Felder:2000hq} in a quartic potential ${1\over 4} \lambda_\phi \phi^4$ and $z$ is defined by $dz = \sqrt{\lambda_\phi} \phi_0 \, dt / a(t)$.
{\it Right:} inflaton--Higgs thermalization constraint.
}
\end{figure}

The specifics of thermalization depend on the momentum distribution and density of the scalars. After inflation, the Higgses as well as the inflaton fluctuations are produced non--perturbatively via parametric resonance.
It is very efficient in the regime $\lambda_{\phi h} \gtrsim \lambda_\phi$ in the (locally) quartic inflaton potential~\cite{Lebedev:2021tas}, which we will assume in our example.
To account for backreaction effects and rescattering, one normally resorts to lattice simulations. A typical example is shown in Fig.~\ref{therm-fig} (left panel), which displays the energy fraction
as a function of time. One observes that, for this parameter choice, about 20\% of the inflaton energy  gets transferred to the Higgs field by the end of preheating. The process is impeded by the Higgs
self--interaction which creates a large  effective mass term $\lambda_h \langle h^2 \rangle$, where $\langle h^2 \rangle$ is the Higgs field variance \cite{Prokopec:1996rr}.
Towards the end of the simulation, the evolution becomes very slow and proper thermalization cannot be observed.
At this stage, the coherent inflaton background is essentially absent and the system consists of the Higgs and inflaton quanta with some non-thermal momentum distribution.

Motivated by the simulations of preheating, we can make a number of simplifying assumptions in the calculation of the reaction rate.
Since the initial number density of $\phi$ is large, let us approximate $n_\phi$ by the corresponding thermal number density at temperature $T$.
 On the other hand, the Higgs number density is relatively low initially, so one may neglect it. Then the relevant reaction rate of $\phi \phi \rightarrow h_i h_i $ is analogous to that   for freeze--in production in the $relativistic$ regime \cite{Lebedev:2019ton}.
 If the Higgs sector populates faster than the Hubble expansion, one expects thermalization.
Using the full Bose--Einstein distribution for the inflaton quanta, one finds \cite{Lebedev:2019ton},
 \begin{multline}
  \Gamma (\phi \phi \rightarrow h_i h_i )= 4\times  {1\over 2! 2!} \; { \lambda_{\phi h}^2 T \over 16 \pi^5} \\
  \times  \int_{m_\phi}^\infty dE ~E \sqrt{E^2-m_h^2} \int_0^\infty d\eta {    \sinh \eta \over e^{{2E\over T} \cosh\eta }-1}~
\ln {  \sinh    {E\cosh\eta + \sqrt{E^2 -m_\phi^2} \sinh\eta \over 2T}    \over
\sinh   {E\cosh\eta - \sqrt{E^2 -m_\phi^2} \sinh\eta \over 2T}  } \;,
\end{multline}
where $E $ is   half the center--of--mass energy.
To be as explicit as possible,
we have factored out the symmetry factor $1/2!2!$ stemming from 2 identical particles in the initial and final states
as well as a factor of 4 representing 4 Higgs d.o.f. in the symmetric phase.
 At high temperature, the mass parameters in this expression should generally include thermal corrections,
 \begin{eqnarray}
&& m_h^2 \rightarrow m_{h}^2\Bigl\vert_{v=0} +  \left( {3\over 16}g^2  + {1\over 16}g'^2  + {1\over 4} y_t^2
+ {1\over 2} \lambda_h  \right) T^2 \;, \\
&& m_\phi^2 \rightarrow  m_{\phi }^2 + \left(   {1\over 4} \lambda_\phi  + {1\over 6} \lambda_{\phi h} \right) T^2 \;,
\end{eqnarray}
where $v = \langle h \rangle$ is the Higgs VEV; $g,g^\prime, y_t$ are the electroweak gauge couplings and the top quark Yukawa coupling, respectively.
We neglect complications associated with the Higgs mass variation at the electroweak crossover, which makes an insignificant impact on the results.

 To find the lower bound on the coupling required by thermalization, we use the following procedure.
For fixed couplings and zero--temperature masses, we maximize the ratio
 \begin{equation}
 { 2 \Gamma(\phi \phi \rightarrow h_i h_i )  \over 3n_\phi H   } \rightarrow {\rm max}
 \end{equation}
with respect to $T$.
If the ratio exceeds one, thermalization is said to occur or, more precisely,  the necessary condition for thermalization has been fulfilled.
In the above expression,
 $n_\phi$ is computed with the full Bose--Einstein distribution and $H$ is given by
 \begin{equation}
  H=\sqrt{\frac{\pi^2g_*}{90}} \, {T^2\over M_{\rm Pl}}   ~~,~~ g_*=1 \;,
  \label{HR}
\end{equation}
 which assumes that the energy density is dominated by a thermal bath of $\phi$.

Our result is presented in Fig.~\ref{therm-fig}, right panel. We find that, for large $m_\phi$, the above ratio is maximized at $T \sim m_\phi$, while for $m_\phi \ll m_h$,
it reaches its maximum at $T\sim m_h$. The thermal mass corrections do not play a significant role in this case.
These results are largely consistent with those obtained by a somewhat different method in \cite{DeRomeri:2020wng}, taking into account the difference in the active d.o.f. $g_*$.
In particular, at $m_\phi \gg m_h$, thermalization requires
\begin{equation}
\lambda_{\phi h} \gtrsim 4 \times 10^{-8} \; \sqrt{m_\phi / {\rm GeV}} \;.
\end{equation}

 It should be noted that the derived bound is subject to some uncertainty stemming from a non--thermal momentum distribution for the inflaton field, neglected Bose--Einstein enhancement  for the Higgses
 as well as a variation of $g_*$ during thermalization. While these factors  introduce ${\cal O}(1)$ uncertainty in the bound, they are not expected to affect the results significantly.

\subsection{\texorpdfstring{$\phi \phi \rightarrow h$}{phi phi -> h}}

At temperatures below the electroweak crossover critical temperature $T_c$, the Higgs field develops a non--zero VEV. This generates the interaction term ${v\over 2 } \lambda_{\phi h} h \phi^2$, which
allows for the fusion reaction  $\phi \phi \rightarrow h$  if $m_\phi < m_h/2$. The corresponding reaction rate per unit volume is  \cite{DeRomeri:2020wng}
 \begin{equation}
\Gamma_{\phi\phi \rightarrow h} = {\lambda_{\phi h}^2 v^2 m_h T \over 32 \pi^3  }\; \theta(m_h- 2m_\phi) \int_0^\infty  d\eta {\sinh\eta \over e^{m_h \cosh \eta \over T}-1} \;
\ln {   \sinh { m_h \cosh \eta +  \sqrt{m_h^2-4m_\phi^2}  \sinh\eta  \over 4T}  \over
 \sinh { m_h \cosh \eta -  \sqrt{m_h^2-4m_\phi^2}  \sinh\eta   \over 4T}}  \;.
\end{equation}
This expression is valid for a single Higgs d.o.f. We neglect the gauge boson contribution at low energies.

The fusion reaction is efficient  in a relatively narrow temperature range: $T$ has to be below the critical temperature, but not much below the Higgs mass. On the other hand,
its rate is enhanced by the phase space factor compared to the $2\rightarrow 2$ reaction rate.
Since $\phi \phi \rightarrow h$ is operative at $T \lesssim T_c$, one needs to account for the Higgs mass and VEV variation around the critical temperature.
 Motivated by  the lattice study \cite{DOnofrio:2015gop}, we
take $T_c \simeq 162$ GeV and
parametrize the Higgs VEV by
$v(T)   =  \alpha\, T \sqrt{162 {\rm \, GeV}-T} $ for $96 \;{\rm GeV }<T<162 \; {\rm GeV}$ and $v(T)=246$ GeV for $T< 96 \; {\rm GeV}$, where $\alpha$ is a fitted constant.
Similarly, we parametrize the Higgs mass as $m_h (T)= \beta \; T (175 {\rm \, GeV}-T)$ for $96 \;{\rm GeV }<T<162 \; {\rm GeV} $ and $m_h (T) =125 $ GeV for
$T< 96 \; {\rm GeV}$, where $\beta$ is another fitted constant.
We then maximize the ratio
\begin{equation}
 { 2 \Gamma(\phi \phi \rightarrow h)  \over 3n_\phi H   } \rightarrow {\rm max}
 \end{equation}
with respect to $T$ in the allowed range and derive the lower bound on $\lambda_{\phi h } $, requiring that this ratio be greater than one.

The resulting thermalization bound is displayed in Fig.~\ref{therm-fig} (right panel) at $m_\phi < m_h/2$. We observe that, in this mass range, the fusion channel gives the dominant contribution
to the Higgs production rate.

\section{Inflaton freeze--out}

One of the challenges for cosmological model building is to suppress the relic density of dark matter. Within the WIMP paradigm, this is achieved via efficient DM annihilation.
In this work, we study a different possibility: a thermal inflaton itself undergoes  annihilation in the SM thermal bath.
This reduces its energy density contribution and if it is the sole source of dark matter, the relic density of the latter will automatically be suppressed.
In our scenario, the inflaton decays into DM after freeze--out, so its decay width has to be sufficiently small.

 \begin{figure}[t]
\centering{
\includegraphics[scale=0.63]{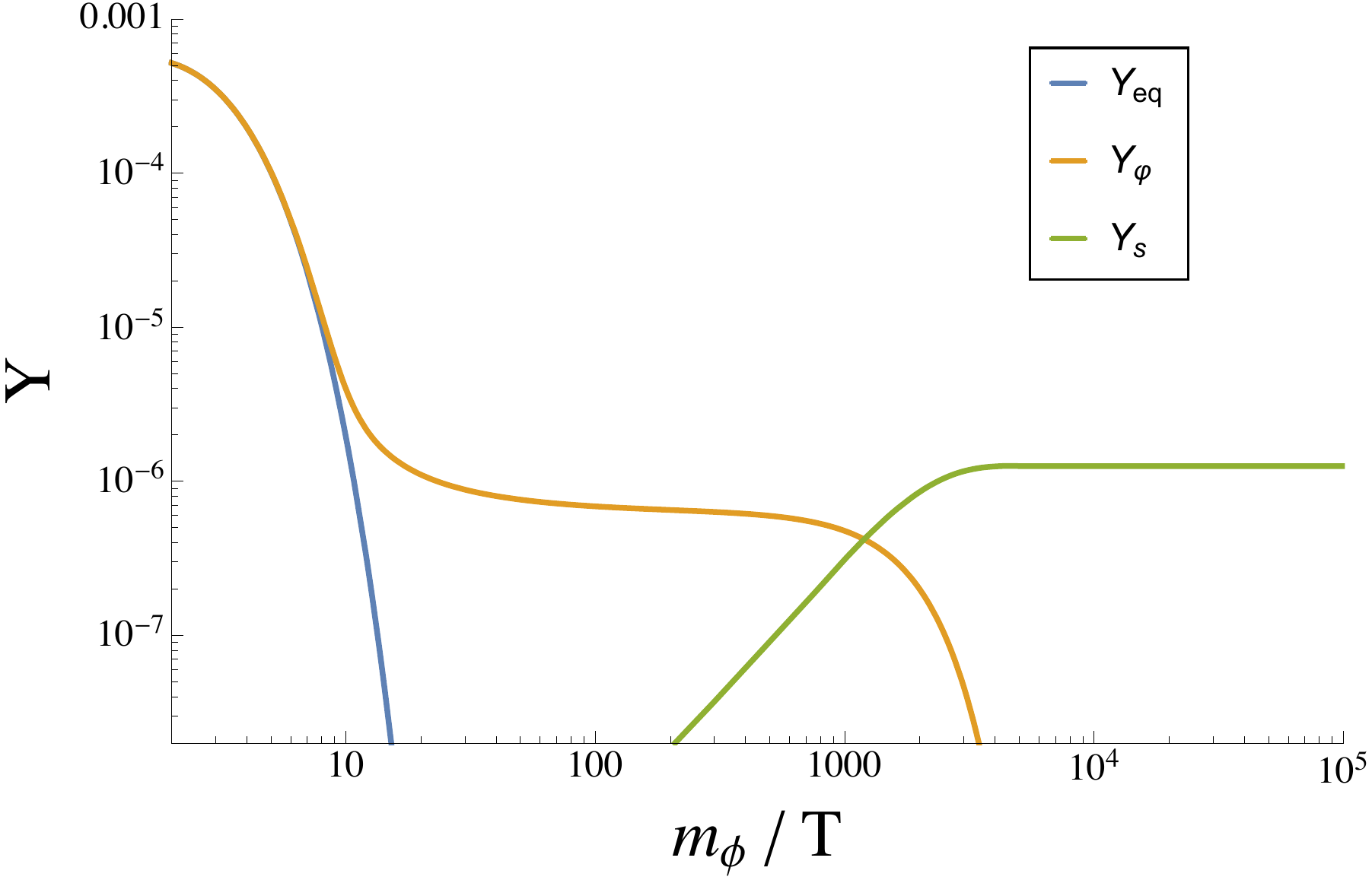}
}
\caption{ \label{FO}
 Example of the inflaton and DM abundance evolution. Here BR$(\phi \rightarrow ss)=1$.}
\end{figure}

Consider inflaton freeze--out in the non--relativistic regime. In this case, one can use the Maxwell--Boltzmann distribution function and the reaction rates take a simple form.
Neglecting the inflaton contribution to the total energy density at this stage, we have
\begin{eqnarray}
\dot n_\phi + 3Hn_\phi &=& 2 \langle \sigma (\phi \phi \rightarrow {\rm SM})  {\rm v} \rangle \,  (n^2_{\phi \; {\rm eq}} - n^2_\phi) - \Gamma_\phi \, (n_\phi - n_{\phi \; {\rm eq}} ) \, \nonumber \\
\dot n_s + 3Hn_s &=& 2\Gamma (\phi \rightarrow ss)\, n_\phi \;,
\end{eqnarray}
where $\Gamma_\phi$ is the total inflaton decay width, $ \sigma (\phi \phi \rightarrow {\rm SM}) $ is the inflaton annihilation cross section and $n_{\phi \; {\rm eq}}$ is the equilibrium
number density. In our convention, $\langle \sigma (\phi \phi \rightarrow {\rm SM})  {\rm v} \rangle $ includes a symmetry factor of 1/2 to account for identical particles in the initial state,
so it appears with a factor of 2 in the Boltzmann equation due to a particle number change by 2 units.

An example of the solution in terms of
\begin{equation}
Y_i = {n_i \over s_{\rm SM}}\;,
\end{equation}
where $s_{\rm SM} $ is the SM entropy density, is shown in Fig.~\ref{FO}.
Freeze--out occurs when $n_\phi$ starts deviating from its equilibrium value, i.e. around $m_\phi /T \sim 10$ for this parameter choice, leading subsequently to $n_\phi \gg n_{\phi \; {\rm eq}}$.
After freeze--out, $Y_\phi$ remains approximately constant until
the Hubble rate becomes comparable to $\Gamma_\phi$. Indeed, the annihilation term in the Boltzmann equation can be neglected since it scales as $a^{-6}$ with the scale factor,
while the decay term and the Hubble term scale as $a^{-3}$ and $a^{-5}$, respectively. At some point, $\Gamma_\phi \sim H$ and the inflaton decays quickly producing pairs of DM particles.
The resulting DM relic density fits observations for a wide range of the input parameters, while the DM mass
is restricted to the range   between $m_\phi /2$  and about 10 keV as required by the structure formation constraints. Its couplings to itself and other fields are assumed to be feeble, such that it does not thermalize
nor entail observable signatures.

The relic density can be determined as follows. First, the
 inflaton annihilation rate is computed precisely with micrOMEGAs. At large masses, it is dominated by $\phi \phi \rightarrow h_i h_i$ such that
 $\langle \sigma (\phi \phi \rightarrow h_i h_i) {\rm v} \rangle =  \lambda_{\phi h}^2/ ( 32 \pi m_\phi^2) $. For lower inflaton masses, other channels must also be taken into account.
 The calculation is similar to that for the  ``singlet scalar dark matter'' \cite{Silveira:1985rk,McDonald:1993ex,Burgess:2000yq}, except the relic density is different.

The frozen--out inflaton quanta decay subsequently into dark matter, at least in part.
The  decay rate into the DM states is given by
\begin{equation}
\Gamma (\phi \rightarrow ss) = {\sigma_{\phi s}^2 \over 32 \pi m_\phi} \;,
\end{equation}
assuming $m_\phi \gg m_s$. If the inflaton is sufficiently heavy, it can also decay directly into the Higgs pairs, $ \Gamma (\phi \rightarrow h_i h_i) = {\sigma_{\phi h}^2 \over 8 \pi m_\phi}$,
where 4 Higgs d.o.f. have been included. Otherwise, it decays into lighter SM states at 1--loop. Depending on $\sigma_{\phi h}/ \sigma_{\phi s}$, this channel may be significant or
suppressed. For our purposes, the results are conveniently parametrized in terms of
the branching ratio
BR$(\phi \rightarrow ss) = \Gamma (\phi \rightarrow ss) / ( \Gamma (\phi \rightarrow ss) +
\Gamma (\phi \rightarrow {\rm SM}) )$.
The inflaton energy and number densities after freeze--out are small compared to those of the SM states, so the entropy and energy injection resulting from its decay can be neglected.
Since the SM sector entropy is conserved,
\begin{equation}
Y_s = Y_\phi^{\rm FO} \times 2 \,{\rm BR} (\phi \rightarrow ss) \;,
\label{Ys}
\end{equation}
   where $Y_\phi^{\rm FO} $ is the inflaton abundance after freeze--out.
 Imposing the observational constraint on $Y_s$ and parametrizing $Y_\phi^{\rm FO} $  in terms of $R$ as
 \begin{equation}
Y_s = 4.4 \times 10^{-10}\; {{\rm GeV} \over m_s} ~~,~~ Y_\phi^{\rm FO} = 4.4 \times 10^{-10}\; {{\rm GeV} \over m_\phi} \; \times R \;,
\end{equation}
we have
\begin{equation}
{m_\phi \over R} = 2 m_s \; {\rm BR} (\phi \rightarrow ss)\;.
\label{solution}
\end{equation}
If $R=1$, the energy density of the inflaton after freeze--out would match that required of DM, so if it {\it were stable}, it would be a good DM candidate.
  Since the branching ratio is bounded from above by 1 and $m_\phi > 2 m_s$ as required by kinematics,
  \begin{equation}
R \geq 1\;.
\end{equation}
For $R>1$, the correct DM density imposes a constraint on a product of $m_s$ and BR$(\phi \rightarrow ss)$, which leads to a one--parameter family of solutions.

\begin{figure}[t]
\centering{
\includegraphics[scale=0.525]{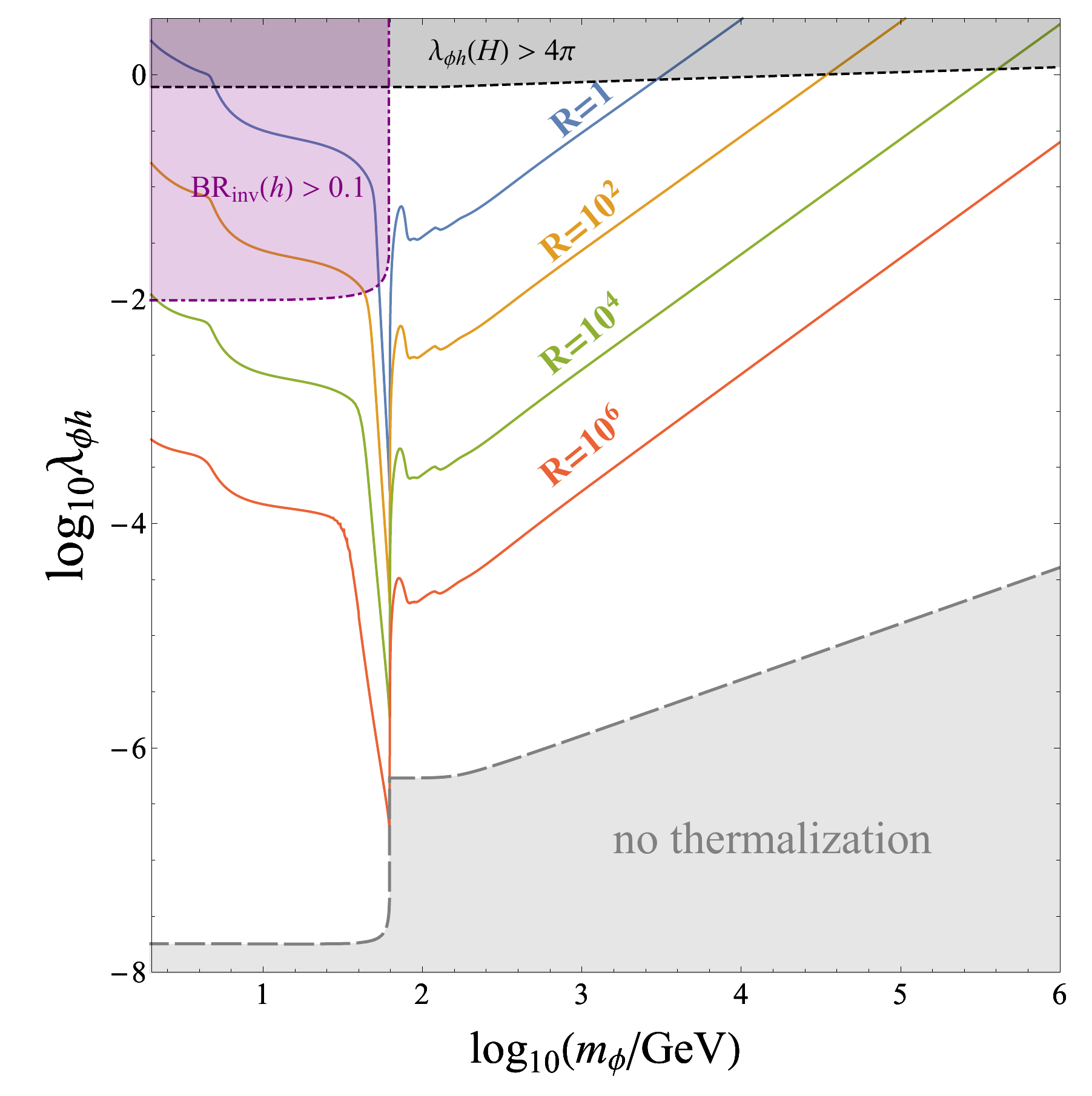}
}
\caption{ \label{main-fig}
 $\lambda_{\phi h}$ vs $m_\phi$ reproducing the correct DM relic density for fixed $R=1,10^2,10^4,10^6$. The shaded areas are excluded by perturbativity at the inflation scale ($H=10^{14}$ GeV),
 invisible Higgs decay and thermalization through $\phi \phi \rightarrow h_i h_i \;, \; \phi \phi \rightarrow h$. The dark matter mass is determined by Eq.~\eqref{solution}.}
\end{figure}

The results  are conveniently presented in terms of $\lambda_{\phi h}, m_\phi$ and $R$. Once these are fixed, the dark matter mass and BR$(\phi \rightarrow ss)$
are determined by Eq.~\eqref{solution}.
Fig.~\ref{main-fig} displays our numerical results  produced with the help of micrOMEGAs \cite{Belanger:2008sj}. The curves with fixed $R$ exhibit non--relativistic inflaton freeze--out, leading
to the correct DM relic density  via~\eqref{solution}. The large $m_\phi$ behavior can be readily understood: the dominant annihilation channel is $\phi \phi \rightarrow h_i h_i$ and
\begin{equation}
\lambda_{\phi h} \simeq  3 \times 10^{-4} ~{m_\phi/{\rm GeV} \over \sqrt{R}} \;,
\end{equation}
where the logarithmic dependence on $\lambda_{\phi h}$ has been neglected.
At low masses, other channels become important, while at $m_\phi \simeq m_h/2$ the annihilation becomes resonantly enhanced.

Close to the resonant annihilation region, kinetic equilibrium  may be lost before inflaton freeze--out, which leads to complications in precision calculations of the relic density \cite{Binder:2017rgn}. This is because
the elastic scattering rate scales as $n_\phi n_{\rm SM} \langle \sigma_{\rm el} {\rm v} \rangle$, while the annihilation rate is proportional to $n_\phi^2   \langle \sigma_{\rm ann} {\rm v} \rangle$.
Away from the resonance, the elastic ($\sigma_{\rm el}$) and annihilation  ($ \sigma_{\rm ann}$) cross sections are not vastly different, while the annihilation rate suffers from an
additional suppression factor $n_\phi$. Thus, kinetic equilibrium is normally maintained at freeze--out, apart from the resonant annihilation region where $ \sigma_{\rm ann} \gg \sigma_{\rm el}$.
The above complication, however, does not make a tangible impact at the level of our precision.

\subsection{Constraints}

 The parameter space is subject to the following constraints.
First of all, the Higgs--inflaton system does not reach thermal equilibrium unless the coupling
is above the bound shown in Fig.~\ref{therm-fig} (right panel).

Further, our framework includes inflation and as such must be perturbative from low energies to
at least the Hubble scale $H$. The couplings at high energies are found via renormalization group (RG) equations \cite{Lebedev:2021xey}:
\begingroup
\allowdisplaybreaks
\begin{eqnarray}
16\pi^2 {\frac{d\lambda_h}{dt}}
&=&
24 \lambda_h^2 -6 y_t^4 + \frac{3}{8}
\left( 2 g^4 + (g^2 + g^{\prime 2})^2 \right)
+ (12 y_t^2 -9 g^2 -3 g^{\prime 2}) \lambda_h + {1\over 2} \lambda_{\phi h}^2 \;,
\nonumber \\
16\pi^2 \frac{d\lambda_{\phi h}}{dt}
&=&
4 \lambda_{\phi h}^2 + 12 \lambda_h \lambda_{\phi h}
-\frac{3}{2} (3 g^2 + g^{\prime 2}) \lambda_{\phi h}
+ 6 y_t^2 \lambda_{\phi h} + 6 \lambda_\phi \lambda_{\phi h} \;,
 \nonumber \\
 16\pi^2 \frac{d \lambda_\phi}{dt}
 &= &
 2 \lambda_{\phi h}^2 + 18 \lambda_\phi^2 \;.
\end{eqnarray}%
\endgroup
Here $t = \ln \mu$ with $\mu$ being the RG scale; $g,g^\prime, y_t$ are the electroweak gauge couplings and the top quark Yukawa coupling, respectively, which run according to the SM RG equations.
Requiring that all the couplings remain
perturbative, $\lambda_i < 4\pi$, between $\mu \sim m_\phi$, where inflaton annihilation occurs,  and $H\sim 10^{14}$ GeV, we find that the values $\lambda_{\phi h}  (m_\phi) \gtrsim 1$ are excluded as shown in Fig.~\ref{main-fig}. (We conservatively assume $\lambda_\phi  (m_\phi) \sim 0$.)

Another constraint is imposed by the Higgs invisible decays. If the inflaton is light, $h \rightarrow \phi \phi$ is kinematically allowed and
\begin{eqnarray}
\Gamma_{h\rightarrow \phi \phi}=\frac{\lambda_{\phi h}^2 v^2}{32\pi m_{h}} \sqrt{1-{\left(\frac{2m_{\phi}}{m_{h}}\right)}^2}  \; \theta(m_{h}-2m_{\phi}) \;.
\end{eqnarray}
Requiring the invisible decay branching ratio to be below 0.1 (see e.g. \cite{Arcadi:2021mag})
rules out significant $\lambda_{\phi h} > 10^{-2}$ for a light inflaton (Fig.~\ref{main-fig}, purple region).

We note that our non--relativistic inflaton freeze--out approximation breaks down at very large $R \sim 10^8 $ since $m_\phi / T_{\rm FO}$ decreases steadily  with growing $R$.
Although this does not rule out the corresponding parameter space, a more careful treatment with the full Bose--Einstein
distribution function  would be necessary in this case.

Finally, there are constraints on the lifetime of the inflaton.  Late inflaton decay into the SM states can spoil the standard nucleosynthesis. Therefore, if the branching ratio BR$(\phi \rightarrow {\rm SM})$
is non--negligible, one would require
\begin{equation}
\tau_\phi < 0.1 \; {\rm sec} \;.
\end{equation}
Parametrizing the total width in terms of ${\rm BR}(\phi \rightarrow ss)$ and $\sigma_{\phi s}$, we thus find
\begin{equation}
\sigma_{\phi s} > 3 \times 10^{-11}\, \sqrt{ {\rm BR}(\phi \rightarrow ss) \;m_\phi \; {\rm GeV}} \;.
\label{BBN}
\end{equation}
In the limit $\sigma_{\phi h} \rightarrow 0$, this constraint is {\it lifted}. For a 100~GeV
inflaton, the typical lower bound on $\sigma_{\phi s}$ is in the sub-eV region.

There is an additional constraint on $\sigma_{\phi s}$ stemming from structure formation considerations. At the time of structure formation, which can be taken to be of order
$\mathcal{O}(1)$~keV, dark matter should be non--relativistic. If the inflaton decays at
temperature $T_{\rm dec}$, the $s$--quanta become nonrelativistic at the temperature
$\sim T_{\rm dec} \times m_s/m_\phi$. Requiring this temperature to be greater than
$1$~keV and eliminating $m_s$ in favor of $R$ results in the constraint
\begin{equation}\label{str-form}
  \sigma_{\phi s} > 10^{-13} \times {\rm BR}(\phi\rightarrow ss)^{3/2 }\,R\,\sqrt{m_\phi\,{\rm GeV}}\,.
\end{equation}
For $ R \times {\rm BR}(\phi \rightarrow ss) > 10^2$ this bound is stronger than (\ref{BBN}), while at $ {\rm BR}(\phi \rightarrow ss) \ll 1$ it becomes weak.

The upper bound on $\sigma_{\phi s}$ is imposed by requiring that the inflaton decay occurs after freeze--out. Comparing the corresponding Hubble rates and ignoring a logarithmic $\lambda_{\phi h}$--dependence,  one finds
\begin{equation}\label{sigma-upper-bound}
  \sigma_{\phi s} \lesssim {\cal O}(1) \times m_\phi \; \left({m_\phi \, {\rm BR}(\phi \rightarrow ss) \over M_{\rm Pl}}\right)^{1/2} \;.
\end{equation}
Thus, the trilinear coupling is suppressed by the factor $(m_\phi / M_{\rm Pl})^{1/2}$ relative to the inflaton mass. For $\sigma_{\phi h}$ of the same order or smaller, the consequent Higgs--inflaton mixing
\cite{Ema:2017ckf} is unobservably small. We note that this bound is easily compatible with (\ref{str-form}) at $ {\rm BR}(\phi \rightarrow ss) \ll 1$.

We note that the coupling pattern discussed in this work is technically natural, i.e. stable under radiative corrections. The trilinear couplings $\sigma_{\phi h}$ and $\sigma_{\phi s}$ break the
$\phi \rightarrow -\phi$ symmetry and their beta--functions are proportional to the tree level values of these couplings (see e.g.\,\cite{Lebedev:2021xey}), hence they can be chosen small. The beta functions for $\lambda_{\phi h}$ and  $\lambda_{\phi s}$
are also proportional to  the couplings themselves. Finally, the Higgs portal coupling $s^2 h^2$ is generated by integrating out the inflaton, however the result
is suppressed by $\sigma_{\phi h} \sigma_{\phi s} /m_\phi^2$, which makes it completely negligible in view of (\ref{sigma-upper-bound}).

In this work, we are {\it assuming} that other sources of dark matter production are subleading. In particular, DM can  be produced during inflation via scalar fluctuations, yet
its abundance can be reduced by introducing a small self--coupling~\cite{Markkanen:2018gcw}. Dim--6 Planck suppressed operators such as $\phi^4 s^2/M_{\rm Pl}^2$  can be
very efficient during preheating~\cite{Lebedev:2022ljz}, however their Wilson coefficients could only be evaluated within a UV complete theory of quantum gravity.
In our phenomenological approach,  we neglect this contribution, which introduces uncertainty in our calculation.

We conclude that inflaton freeze--out is an efficient mechanism for suppressing the DM relic density that allows  for a vast range of the inflaton and DM masses.
In the extreme case, the inflaton itself could be identified with dark matter, yet the direct detection constraint is then very strict and forces $m_\phi$ into a very narrow range
close to $m_h/2$ \cite{Lebedev:2021zdh}.\footnote{This conclusion does not apply to non--minimal inflaton DM models \cite{Garcia:2021gsy}.} In our scenario, however, the direct detection bound is irrelevant\footnote{The indirect DM detection signal is also highly suppressed: DM annihilation proceeds via the inflaton such that the amplitude contains the factor $\sigma_{\phi h} \sigma_{\phi s} /m_\phi^2$, leading
to the cross section suppressed by $m_\phi^2 /M_{\rm Pl}^2 $ in view  of (\ref{sigma-upper-bound}). This makes dark matter virtually undetectable.}
and the inflaton energy density can exceed that of dark matter by orders of magnitude,
which widens significantly available parameter space.

It is interesting that
the inflaton is allowed to  be light and have a substantial  coupling to the Higgs boson, in which case some of the allowed  parameter space can be probed via invisible Higgs decay at the LHC \cite{Djouadi:2011aa}.
Indeed, its high luminosity phase aims at detecting the invisible decay
with a  branching ratio above  2.5\% \cite{Cepeda:2019klc}, which is a twofold improvement in the coupling sensitivity compared to that of Fig.~\ref{main-fig}.
Heavier inflaton pairs could be produced, for example, via vector boson fusion $VV \rightarrow h \rightarrow \phi \phi$ \cite{Djouadi:2011aa}. However, the corresponding cross section
is suppressed either by small couplings or large masses, and the process appears to be beyond the reach of the LHC.

\section{Conclusion}

We have studied the possibility that, after inflation,  the inflaton reaches thermal equilibrium with the SM thermal bath and subsequently freezes--out. After freeze--out, it decays producing non--thermal dark matter.
This mechanism suppresses  the DM relic density without requiring a significant coupling between the Standard Model and dark matter, thereby evading strong direct detection constraints.
On the other hand, the Higgs coupling to the inflaton can be substantial and lead to observable signatures at the LHC.

In this paper, we have focused on the Higgs portal framework which provides us with a minimal and viable setting to implement the inflaton freeze--out idea. If inflation is driven by a non--minimal scalar--curvature
coupling, the loop corrections induced by the Higgs--inflaton coupling do not adversely affect the inflaton potential,
making  the picture  radiatively stable.
We  expect that this approach could also be implemented in more general settings.
 \\ \ \\
 {\bf Acknowledgements.} TS acknowledges  EDUFI  support.
OL thanks   the IN2P3 master project UCMN.

\end{document}